\documentclass[12pt,preprint,showpacs,preprintnumbers,amsmath,amssymb]{revtex4}

\usepackage{graphicx}
\usepackage{dcolumn}
\usepackage{bm}
\usepackage{color}

\begin{document}


\title{Deformed-Basis Antisymmetrized Molecular Dynamics\\
and Its Application to $^{\bf 20}{\bf Ne}$}

\author{Masaaki Kimura}
\affiliation{RI-Beam Science Laboratory, RIKEN (The Institute of
Physical and Chemical Research), Wako, Saitama 351-0198, Japan.}


\date{\today}

\begin{abstract}
A new theoretical framework is presented that uses a localized triaxially
deformed Gaussian as the single-particle wave packet. The model space
enables us to adequately describe both the deformed mean-field
structure and the cluster structure within the same  framework. The
improvement over the original version of antisymmetrized molecular
dynamics which uses the spherical Gaussian is verified by its
application to the $^{20}{\rm Ne}$ nucleus. The coexistence and
interplay of the cluster structure and the deformed mean-field
structure in the low-excited states of $^{20}{\rm Ne}$ is
studied. In particular, the intra-band $E2$ transition probabilities in
$K^\pi$=$0^+_1$ and $2^-$ bands are reproduced without any effective
charge.   
\end{abstract}

\pacs{Valid PACS appear here}
\maketitle

\section{Introduction}\label{SEC_INTRO}
In light nuclei up to the beginning of the $sd$-shell, the cluster
structure is one of the most essential features of the nuclear structure
together with a mean-field-like (shell-model-like) structure. On the
heels of the cluster structure studies of the $p$-$sd$ shell
nuclei \cite{P_SHELL_CLUSTERS}, many theoretical studies have been conducted
to investigate the cluster aspects of the heavier $sd$-$pf$ shell nuclei
\cite{FRIEDRICH,OHKUBO}. Experimentally, some negative-parity levels
which have large $\alpha$-spectroscopic factors have been observed in
$^{40}{\rm Ca}$ and $^{44}{\rm Ti}$ \cite{CLUSTER_Ti},
and these states have been considered to be $K^\pi$=$0^-$ band members of
the parity doublet bands with $K^\pi$=$0^+$ which have an 
$\alpha$+core structure. It is of great interest and importance to survey 
the cluster aspect of these heavier isotopes, but the study seems to be
in a rather intermediate stage. Another new field for cluster study is
the unstable nucleus. Due to advances in experimental technique, many
new aspects and phenomena such as neutron halo, neutron 
skin and the change of the magic number \cite{NEUTRON_RICH_EXP,OTSUKA}
have been observed and studied, and now an increasing body of 
information on the unstable nuclei is available. It has been shown by
many theoretical and experimental 
studies that the cluster-like correlation is also important and the
exotic cluster structure which cannot be seen in stable isotopes appears
in light neutron-rich isotopes. Therefore, new cluster aspects are
expected in heavier stable and unstable nuclei.   

One of the difficulties of cluster structure study in heavier nuclei
is the existence and importance of the mean-field deformation of
the nucleus, which should be properly treated together with
clustering. It is quite conceivable that the mean-field 
deformation and clustering coexist, compete or are mixed. For
the study of the nucleus in such situations, we have to treat the
cluster structure and the deformed mean-field structure within the same
framework. Another difficulty is the increasing importance of the
spin-orbit force. The strong effect of the spin-orbit force will
dissolve the cluster structure. Therefore the expectation value of the
spin-orbit force can be regarded as a kind of  measure that tells
whether or not a cluster structure with doubly closed shell subunits
develops.  The importance of the spin-orbit force increases as we
proceed to heavier nuclei.    

To study the cluster aspects of heavier nuclei, we have
introduced deformed-basis antisymmetrized 
molecular dynamics (deformed-basis AMD), which employs a localized
triaxially deformed Gaussian as the single-particle wave packet
\cite{KIM}. In this work, we present the detailed framework of 
deformed-basis AMD and argue the advantages of this new framework. In
short, the present framework has two advantages. The first is its better
description of the deformed mean-field achieved by making the
single-particle wave packets deformable. The second is its improved
incorporation of spin-orbit force effects. These advantages become clear
in its application to the low-lying rotational bands of the $^{20}{\rm
Ne}$ nucleus.

The $K^\pi$=$0^\pm_1$ bands of $^{20}{\rm Ne}$ are regarded as 
parity doublet bands with an $\alpha$+$^{16}{\rm O}$ cluster
structure. However, at the same time, the impurity of the
$\alpha$+$^{16}{\rm O}$ cluster structure in the ground band has been
discussed by many authors, while the $K^\pi$=$0^-$ band is considered to
have an almost pure $\alpha$+$^{16}{\rm O}$ cluster structure. Indeed
the intra-band B($E2$) values of the ground band have been
underestimated by $\alpha$+$^{16}{\rm O}$ cluster models. It is
considered that this deficiency originates in the mixture of the
deformed mean-field structure with an $\alpha$+$^{16}{\rm O}$ cluster
structure. The lowest negative-parity band is the $K^\pi$=$2^-$ band
which has a $(0p)^{-1}(sd)^5$ structure. Unlike $K^\pi$=$0^\pm_1$ bands,
this band should be understood in terms of the deformed
mean-field. Thus, in the low-lying states of $^{20}{\rm 
Ne}$, the cluster structure and the deformed mean-field structure
coexist and are mixed with each other. This intermingled situation has not
been treated in a consistent manner within a full microscopic
framework. It will be shown that the present framework is
flexible enough to describe the coexistence and the mixture of the
cluster structure and the deformed mean-field structure.

The contents of this article are as follows. In the next section, the
framework of deformed-basis AMD is given and the calculational
procedure adopted in this study is explained. The features and the
advantages of this new framework are explained in the section
\ref{SEC_FEATURE}. The advantages in the practical calculation are
shown by applying the method to the $^{20}{\rm Ne}$ nucleus
(section \ref{SEC_APPLI}). In the course of this application, the
characters of the low-lying rotational bands of $^{20}{\rm Ne}$,
$K^\pi$=$0^+_1,0^+_4,0^-_1$ and $2^-$ bands are discussed, and the
calculated results are examined. In the last section, we summarize this
work.    

\section{Framework of deformed-basis AMD}\label{SEC_FRAME}
In this section, the framework of deformed-basis AMD is presented and
the calculational procedure adopted in this study is explained. In the
following, we call usual AMD, which uses a spherical Gaussian as the
single-particle wave packet 'spherical-basis AMD' to keep the
distinction clear. For a more detailed explanation of spherical-basis
AMD framework, the reader is directed to references \cite{ONO,ENYO_OLD}.

\subsection{Wave function and Hamiltonian}\label{SUBSEC_WAVEFUNC}
In deformed-basis AMD, the intrinsic wave function of the system with
mass A is given by a Slater determinant of single-particle wave packets;
\begin{eqnarray}
 \Phi_{int} &=& \frac{1}{\sqrt{A!}}{\mathcal A}
  \{\varphi_1,\varphi_2,...,\varphi_A \} ,\label{EQ_INTRINSIC_WF}\\
 \varphi_i({\bf r}) &=& \phi_i({\bf r})\chi_i\xi_i ,
\end{eqnarray}
where $\varphi_i$ is the $i$th single-particle wave packet
consisting of spatial $\phi_i$, spin $\chi_i$ and isospin
$\xi_i$ parts. Deformed-basis AMD employs the triaxially deformed
Gaussian centered at ${\bf Z}_i$ as the spatial part of the
single-particle wave packet, while spherical-basis AMD limits the
Gaussian to spherical shape: 
\begin{eqnarray}
 \phi_i({\bf r}) &\propto& \exp\biggl\{-\sum_{\sigma=x,y,z}\nu_\sigma
  (r_\sigma - {\rm Z}_{i\sigma})^2\biggr\},\nonumber\\
 \chi_i &=& \alpha_i\chi_\uparrow + \beta_i\chi_\downarrow,
  \quad |\alpha_i|^2 + |\beta_i|^2 = 1\nonumber \\
 \xi_i &=& proton \quad {\rm or} \quad neutron. \label{EQ_SINGLE_WF}
\end{eqnarray}
Here, the complex number parameter ${\bf Z}_i$ which represents the center
of the Gaussian in phase space takes an independent value for each
nucleon. The width parameters $\nu_x, \nu_y$ and $\nu_z$ are real number
parameters and take independent values for each direction, but are
common to all nucleons. Spin part $\chi_i$ is parametrized by $\alpha_i$
and $\beta_i$ and isospin part $\xi_i$ is fixed to up (proton) or
down (neutron). ${\bf Z}_i$, $\nu_x, \nu_y, \nu_z$ and $\alpha_i$,
$\beta_i$ are the variational parameters and are optimized by the method of
frictional cooling explained in the next subsection. By using a
deformed Gaussian basis, we can improve the description of deformed
nuclei. The effects resulting from the deformed Gaussian basis
are explained in the section \ref{SEC_FEATURE} and are shown in its
practical application to the $^{20}{\rm Ne}$ (section \ref{SEC_APPLI}). 
As the variational wave function, we employ the parity projected wave
function in the same way as many other AMD studies
\cite{ONO,ENYO_OLD,DOTE},  
\begin{eqnarray}
 \Phi^{\pm} = P^\pm \Phi_{int} = \frac{(1\pm P_x)}{2} \Phi_{int} ,\label{EQ_PARITY_WF}
\end{eqnarray}
here $P_x$ is the parity operator and $\Phi_{int}$ is the intrinsic wave
function given in Eq(\ref{EQ_INTRINSIC_WF}). 

The Hamiltonian used in this study is as follows;
\begin{eqnarray}
\hat{H} = \hat{T} + \hat{V_n} + \hat{V_c} - \hat{T_g} ,
\end{eqnarray}
where $\hat{T}$ and $\hat{T}_g$ are the kinetic energy and the energy of the
center of mass motion, respectively. The expectation value of the
$\hat{T_g}$ is evaluated without any approximation, but the wave
function contains the spurious component of the center-of-mass motion.
Because the wave function of the center-of-mass also has a deformed
Gaussian form, it becomes inseparable when the angular momentum is
projected and/or the multiple Slater determinants are superposed.
prescription should be developed. In this study, we have approximated
its influence by subtracting the expectation value of the center-of-mass
kinetic energy. We have used the Gogny force D1S force as an effective  
nuclear force $\hat{V}_n$. Coulomb force $\hat{V}_c$ is approximated by
the sum of seven Gaussians. 

\subsection{Energy variation, angular momentum projection and generator
  coordinate method}
We perform the variational calculation and optimize the variational
parameters included in the trial wave function Eq(\ref{EQ_PARITY_WF}) to
find the state that minimizes the energy of the system $E^\pm$;
\begin{eqnarray}
 E^\pm=\frac{\langle\Phi^\pm|\hat{\mathcal H}|\Phi^\pm\rangle}
  {\langle\Phi^\pm|\Phi^\pm\rangle},\quad
\hat{\mathcal H} = \hat{H} + \hat{V}_{cnst}. \label{eq::hamiltonian}
\end{eqnarray}
We add the constraint potential $\hat{V}_{cnst}$to the Hamiltonian
$\hat{H}$ to obtain the minimum energy state under the optional
constraint condition. In this study, we constrain matter quadrupole
deformation by employing the potential $\hat{V}_{cnst} =
v_{cnst}(\langle \beta^2 - \beta_0^2)^2$ and we obtain the
optimized wave function $\Phi^\pm (\beta_0)=P^\pm\Phi_{int}(\beta_0)$ as
a function of the deformation parameter $\beta_0$. The evaluation of the
quadrupole deformation parameter $\beta$ is explained in 
reference \cite{DOTE}. At the end of the variational calculation, the
expectation value of $V_{cnst}$ should be zero in principle, and in the
practical calculations we confirm that it is less than 100 eV.  

Energy variation of the AMD wave function is performed by the frictional
cooling method, which is one of the imaginary time development methods.
The reader is directed to references \cite{ONO,ENYO_OLD} for a more
detailed description. The time development equation for the complex
number parameters ${\bf Z}_i, \alpha_i$ and $\beta_i$ is as follows;  
\begin{eqnarray}
 \frac{dX_i}{dt} = \frac{\mu}{\hbar}\frac{\partial}{\partial X_i^*}
  \frac{\langle\Phi^\pm|\hat{\mathcal H}|\Phi^\pm\rangle}
  {\langle\Phi^\pm|\Phi^\pm\rangle},  \quad (i=1,2,...,A)
\end{eqnarray}
and for the real number parameters $\nu_x$, $\nu_y$ and $\nu_z$;
\begin{eqnarray}
 \frac{d\nu_\sigma}{dt} = \frac{\mu^\prime}{\hbar}\frac{\partial}
  {\partial \nu_\sigma}\frac{\langle\Phi^\pm|\hat{\mathcal H}|\Phi^\pm\rangle}
  {\langle\Phi^\pm|\Phi^\pm\rangle},  \quad (\sigma=x,y,z)
\end{eqnarray}
Here $X_i$ is ${\bf Z}_{i}$, $\alpha_i$, or $\beta_i$. $\mu$ and
$\mu^\prime$ are arbitrary negative real numbers. It is easy to show
that the energy of the system decreases as time develops, and after
sufficient time steps we obtain the minimum energy state.

From the optimized wave function, we project out the eigenstate of the
total angular momentum $J$,
\begin{eqnarray}
 \Phi^{J\pm}_{MK}(\beta_0) = P^{J}_{MK}\Phi^{\pm}(\beta_0)
  = P^{J\pm}_{MK}\Phi_{int}(\beta_0).
  \label{EQ_ANGULAR_WF}
\end{eqnarray} 
Here $P^{J}_{MK}$ is the total angular momentum projector. The integrals
over the three Euler angles included in the $P^{J}_{MK}$ are evaluated
by numerical integration.  

Furthermore, we superpose the wave functions $\Phi^{J\pm}_{MK}$ which
have the same parity and angular momentum but have different values of
deformation parameters $\beta_0$ and $K$. Thus the final wave function of
the system becomes as follows:
\begin{eqnarray}
 \Phi_n^{J\pm} = c_n\Phi^{J\pm}_{MK}(\beta_0)
  + c_n^\prime\Phi^{J\pm}_{MK^\prime}(\beta_0^\prime) + \cdots,
  \label{EQ_GCM_WF}
\end{eqnarray}
where quantum numbers other than total angular momentum and parity
are represented by $n$. The coefficients $c_n$, $c'_n$,... are determined
by the Hill-Wheeler equation,
\begin{eqnarray}
 \delta \bigl(\langle\Phi^{J\pm}_n|\hat{H}|\Phi^{J\pm}_n\rangle - 
  \epsilon_n \langle\Phi^{J\pm}_n|\Phi^{J\pm}_n\rangle \bigr) =0.
  \label{EQ_GCM_EQ}
\end{eqnarray}

It is sometime useful to include the intrinsic wave function obtained
from the variational calculation after the projection to the opposite
parity. For instance, suppose that $\Phi_{int(+)}(\beta_0)$ and
$\Phi_{int(-)}(\beta_0)$ are
the intrinsic wave functions obtained by the variational calculation
after the projection to positive- and negative-parity,
respectively. We include $P^{J^+}_{MK}\Phi_{int(-)}(\beta_0)$ in the
GCM basis as well as $P^{J^+}_{MK}\Phi_{int(+)}(\beta_0)$ to describe the
positive-parity state $\Phi_n^{J+}$; 
\begin{eqnarray}
 \Phi_n^{J^+} = c_nP^{J^+}_{MK}\Phi_{int(+)}(\beta_0)
  + c_n^\prime P^{J^+}_{MK^\prime}\Phi_{int(+)}(\beta_0^\prime)
  +\cdots\nonumber\\ 
  + c_n^{\prime\prime} P^{J^+}_{MK^{\prime\prime}}
   \Phi_{int(-)}(\beta_0^{\prime\prime})
   + c_n^{\prime\prime\prime} P^{J^+}_{MK^{\prime\prime\prime}}
   \Phi_{int(-)}(\beta_0^{\prime\prime\prime})+\cdots.
\end{eqnarray}
In the present study, we have employed all of the intrinsic wave
functions on the obtained positive- and negative-parity energy curves
as the GCM basis. 

\section{Features and advantages of the present model}\label{SEC_FEATURE}
The framework of AMD has some advantages over other theoretical
methods. For instance, parity projection before the variation, angular
momentum projection after (before) the variation and superposition of
the Slater determinants (GCM) are easily executable in this
framework. It is well known that parity projection is essentially 
important to describe the parity asymmetric shape of the nucleus and
parity asymmetric cluster structure \cite{ENYO_OLD, YABANA}. Angular
momentum projection and superposition of Slater determinants are
indispensable to study the 
level scheme of the nucleus and the nuclear-structure-change in each
state beyond the mean-field picture. In this section, besides these
advantages which are common to spherical- and deformed-basis AMD, we
explain the features and advantages of deformed-basis AMD in the study
of the heavier nuclei. First, we discuss the description of 
the deformed mean-field structure and cluster structure in this
model. Then we discuss the evaluation of the spin-orbit force.  

\subsection{Deformed mean-field structure and cluster structure}
 In deformed-basis AMD, deformed mean-field structure is
described by deformed single-particle wave packets and the
centroids of single-particle wave packet ${\bf Z}_i$ gathered to the
center of the nucleus, while the cluster structure is described by the
localization of ${\bf Z}_i$s into several parts of the coordinate space.

\begin{figure}
\includegraphics[width=\hsize]{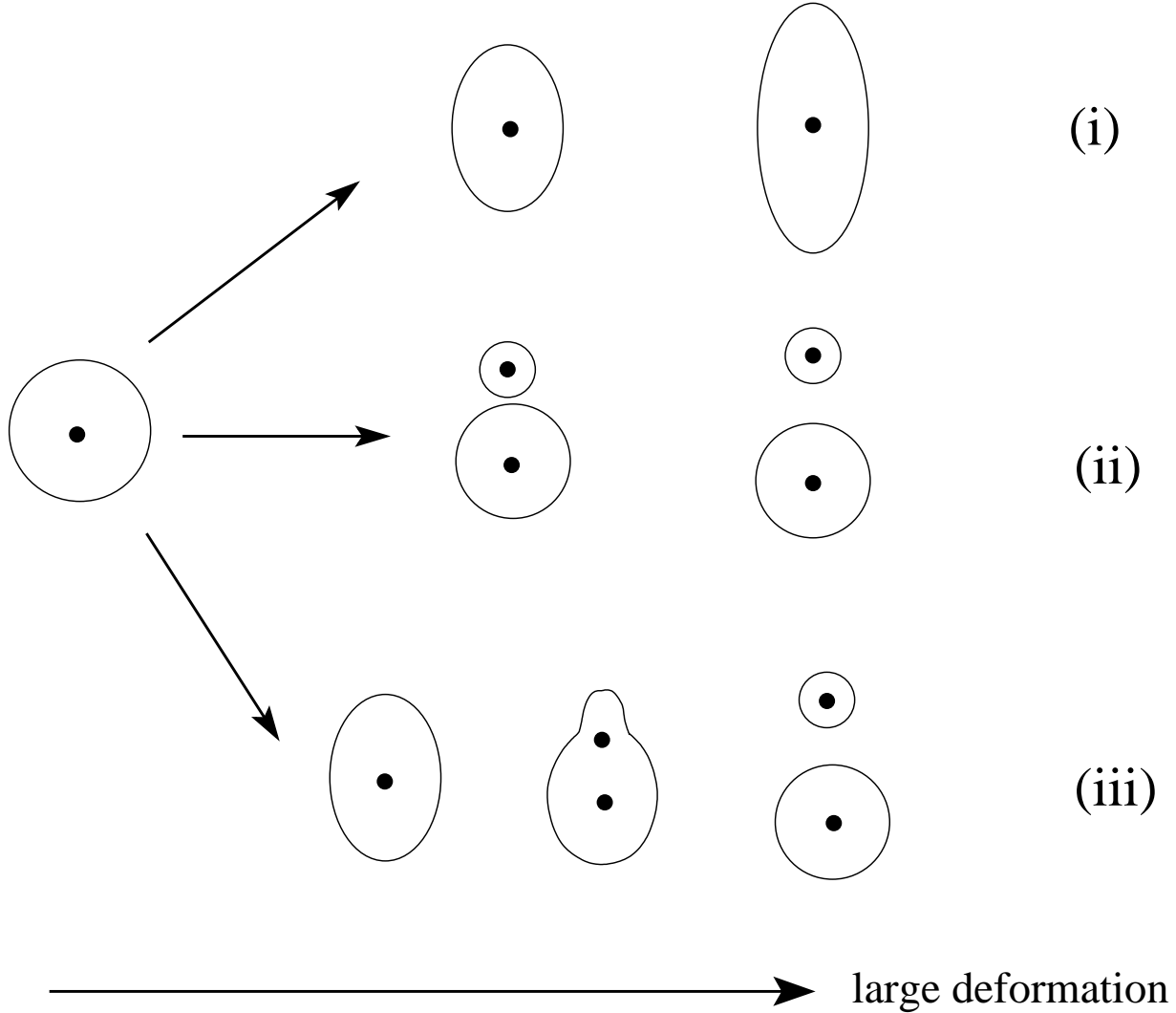}
\caption{Schematic figure showing three different patterns of the
 nuclear deformation in deformed-basis AMD. Points in the figure indicate
 the centroids of the single-particle wave packets (${\rm Re}({\bf Z}_i)$)
 of a deformed-basis AMD wave function.}\label{FIG_ILLUST} 
\end{figure}

Roughly speaking, in deformed-basis AMD, the change of nuclear
shape as a function of nuclear deformation is classified into three
patterns as sketched in FIG. \ref{FIG_ILLUST}. In this figure, for the
sake of simplicity, we consider only the elliptical deformation of the
mean-field and the cluster structure consisting of two spherical
subunits. The first pattern shows deformation of the mean-field
(FIG. \ref{FIG_ILLUST}, (i)). When the nuclear deformation
becomes larger, 
only the deformation of the mean-field becomes larger. In this case, all
${\bf Z}_i$ in Eq(\ref{EQ_SINGLE_WF}) are gathered in the small region
around the center of nucleus and the single-particle wave packets are
deformed. In this limit, the nuclear deformation is described by the
deformation of the single-particle wave packets and the strong effect of
the Pauli principle (single particle motion in the deformed
mean-field). In other words, the intrinsic wave function
Eq. (\ref{EQ_INTRINSIC_WF}) contains a large amount of the deformed 
shell model configurations. This structure is not described well by 
spherical-basis AMD and this is the reason that spherical-basis AMD
sometimes provides an inadequate description of deformed heavier nuclei
\cite{SUGAWA}. The second pattern (FIG. \ref{FIG_ILLUST}, (ii)) shows
the deformation caused by clustering. In this pattern, 
the nucleus splits into two clusters. This structure is described by the
localization of ${\bf Z}_i$ into two parts and the spherical shape of the
single-particle wave packets. As the nuclear deformation becomes larger,
the distance between two clusters becomes larger, but the
single-particle wave packets remain spherical.  The third pattern
(FIG. \ref{FIG_ILLUST}, (iii)) has a mixed character of
(i) and (ii). At small deformations, the nucleus has a mean-field
character, but as the deformation becomes larger, a cluster-like
structure grows and the nucleus has an intermediate character. At large
deformations, it changes to an almost pure cluster structure. In the
application to $^{20}{\rm Ne}$, it will be shown that
$J^\pi$=$2^-$($K^\pi$=$2^-$),$1^-$($K^\pi$=$0^-$) and
$0^+$($K^\pi$=$0^+$) curves correspond to patterns (i),(ii) and (iii),
respectively.   

\subsection{Evaluation of the spin-orbit force}
In medium-heavy nuclei, the importance of the spin-orbit force
increases. In short, the spin-orbit force acts to dissolve the cluster
structure. Since spherical-basis AMD does not assume any cluster
structure, it can describe the dissociation of the cluster structure due
to the spin-orbit force. However, it was pointed out that in heavier
isotopes, the spherical-basis AMD wave function provides a much smaller
expectation value of the spin-orbit force compared to HF theory when
the same interaction Skyrme III force is adopted \cite{SUGAWA}.
We consider that this is because single-particle wave packets are
limited to the 
spherical Gaussian form in spherical-basis AMD. Since the spin-orbit
force contains the derivative operator, it is sensitive to the slope and
the deformation of the single-particle wave packet. When the system favors
the mean-field deformation, doubly closed-shell clusters such as
$\alpha$ and $^{16}{\rm O}$ clusters are difficult to form. This
often means that  deformed-basis AMD gives us larger expectation
values of the spin-orbit force than spherical-basis AMD does, because
the latter tends to give us a larger amount of the clustering component
in the system wave function than the former. For instance,
deformed-basis AMD gives twice as large an expectation value of the
spin-orbit force (-19.0 MeV) as spherical-basis AMD gives (-11.6 MeV) in
the $^{24}{\rm Mg}$ nucleus. In contrast, when a nucleus favors a
cluster structure consisting of doubly-closed-shell subunits like
$K^\pi$=$0^-$ band of $^{20}{\rm Ne}$, spin-orbit force does not act at
all and the shape of the single-particle wave packet is not distorted by
it. Therefore, the expectation value of the spin-orbit force can be
regarded as a type of measure that identifies the formation and
dissociation of the cluster structure. In the next section, it will be
shown that the different characters of the rotational bands of
$^{20}{\rm Ne}$ are directly reflected in the expectation value of the
spin-orbit force.

\section{Application to $^{\bf 20}{\bf Ne}$ and discussion} \label{SEC_APPLI}
We have applied the present framework to the $^{20}{\rm Ne}$ nucleus to
study the coexistence and mixing of the cluster structure and the 
deformed mean-field structure. First, we investigate the structure
change of the wave functions as a function of matter quadrupole
deformation. Then we study the detailed properties of each band. 

\subsection{Overview of the energy curves}
\begin{figure}
\includegraphics[width=\hsize]{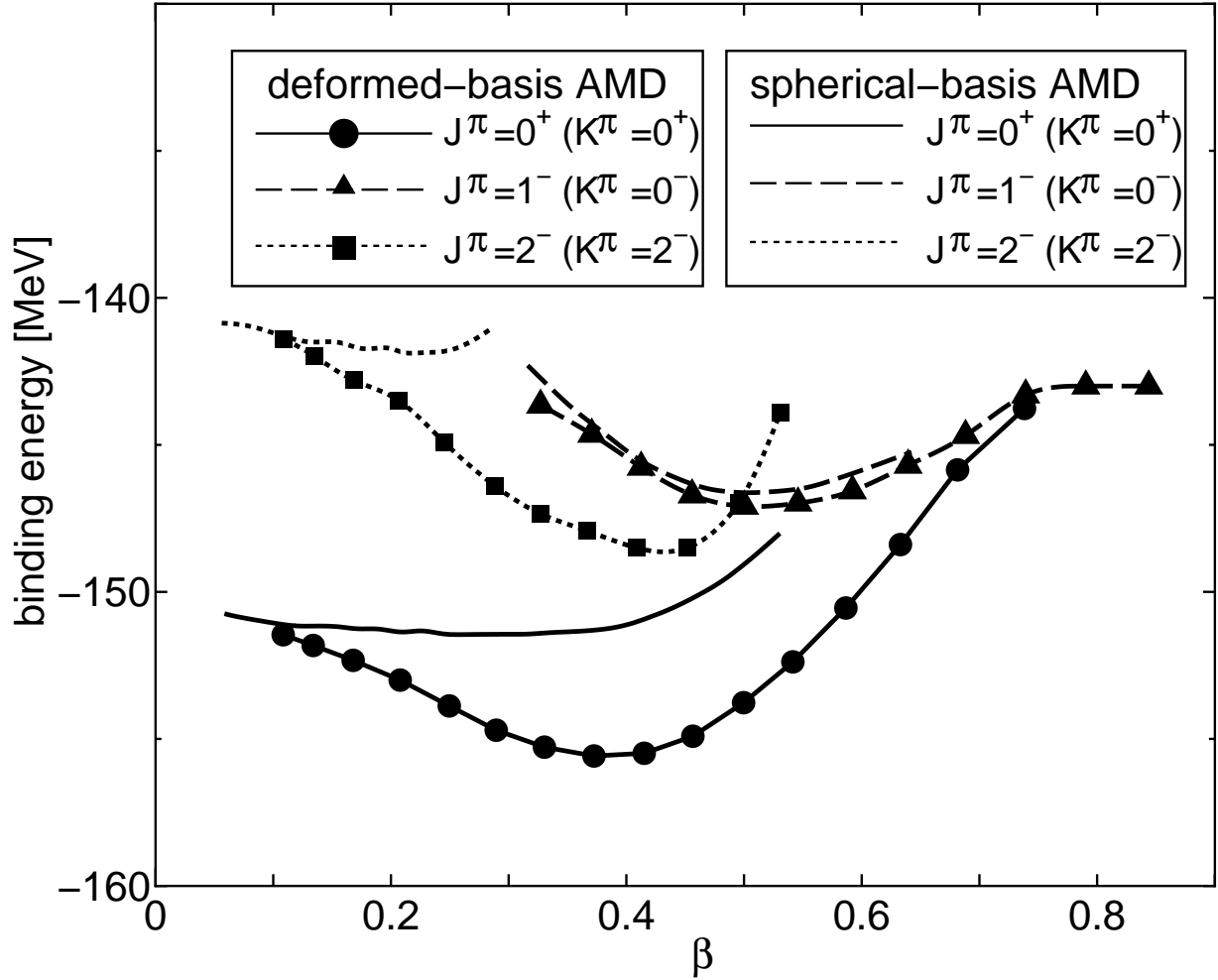}
\caption{$\beta$-energy curves of $^{20}{\rm Ne}$ (energy curves as a
 function of the deformation parameter $\beta$). Lines with symbols
 (Lines) are the result of deformed-basis AMD (spherical-basis AMD).
 $K^\pi$=$0^+$ and $K^\pi$=$2^-$ states are more deeply bound in
 deformed-basis AMD than in spherical-basis AMD (See
 text). }\label{FIG_SURFACE}  
\end{figure}

\begin{figure}
\includegraphics[width=\hsize]{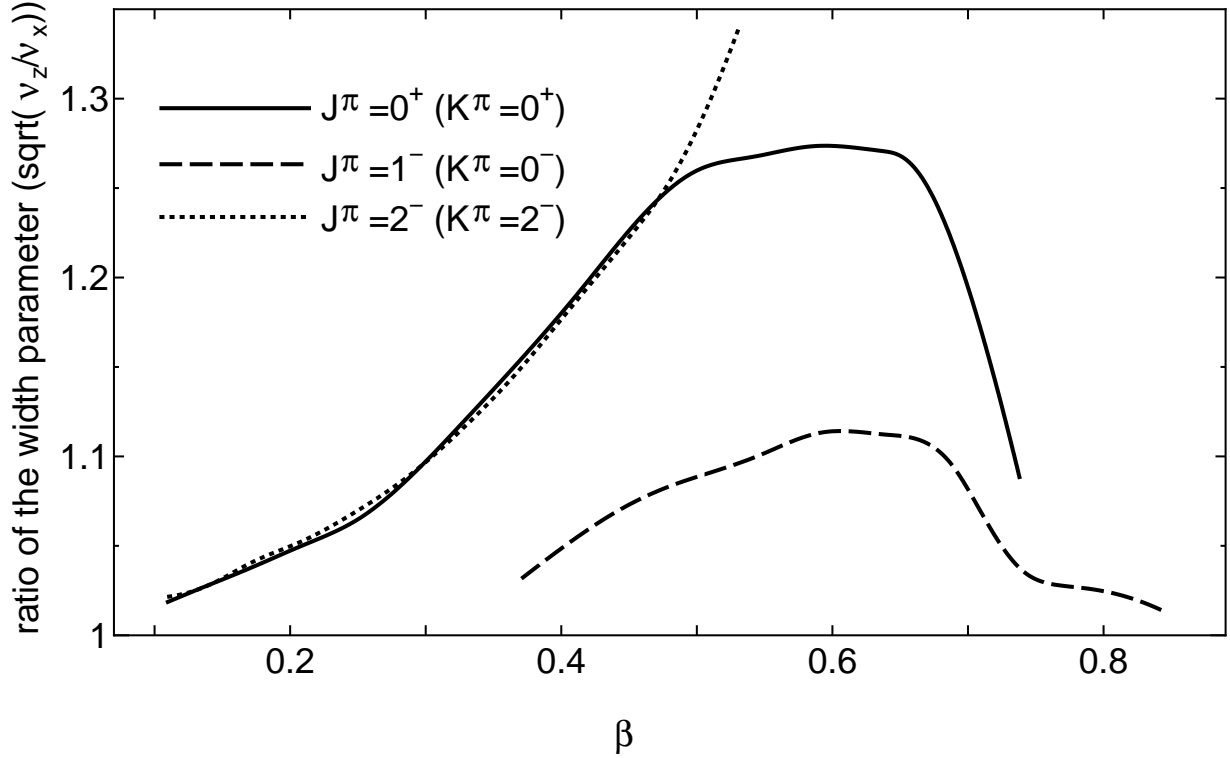}
\caption{The ratio of the width parameter of the single-particle wave
 packet ($\sqrt{{\nu_z}/{\nu_x}}$) as a function of the deformation
 parameter $\beta$. In the present calculation, the single-particle wave
 packets are always almost axially symmetric and $z$ axis is taken to be
 the symmetry axis.}\label{FIG_WIDTH}  
\end{figure}

\begin{figure}
\includegraphics[width=\hsize]{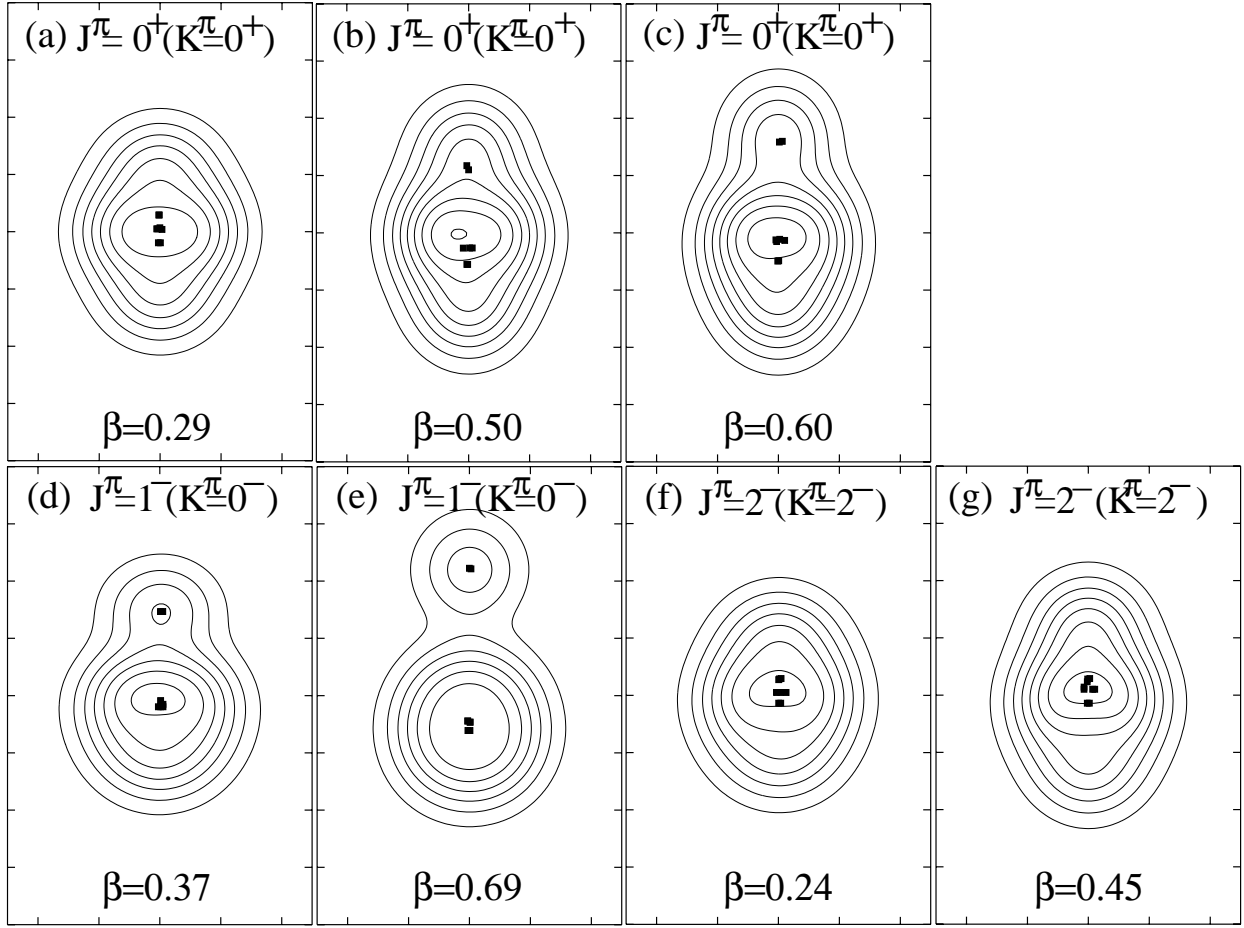}
\caption{The density distributions of the intrinsic wave functions on the
 energy curve obtained by the deformed-basis AMD. Twenty centroids of 
 single-particle wave packets (Re ${\bf Z}_i$, $i$=1$\sim$20) are denoted
 by the black points in each panel. (a), (b) and (c) are for the
 $J^\pi$=$0^+$ $(K^\pi$=$0^+)$ state,  (d) and (e) are for the
 $J^\pi$=$1^-$ $(K^\pi$=$0^-)$ states, and (f) and  (g) are for the
 $J$=$2^-$ $(K^\pi$=$2^-)$ states.}
\label{FIG_Z}
\end{figure}

\begin{figure}
\includegraphics[width=\hsize]{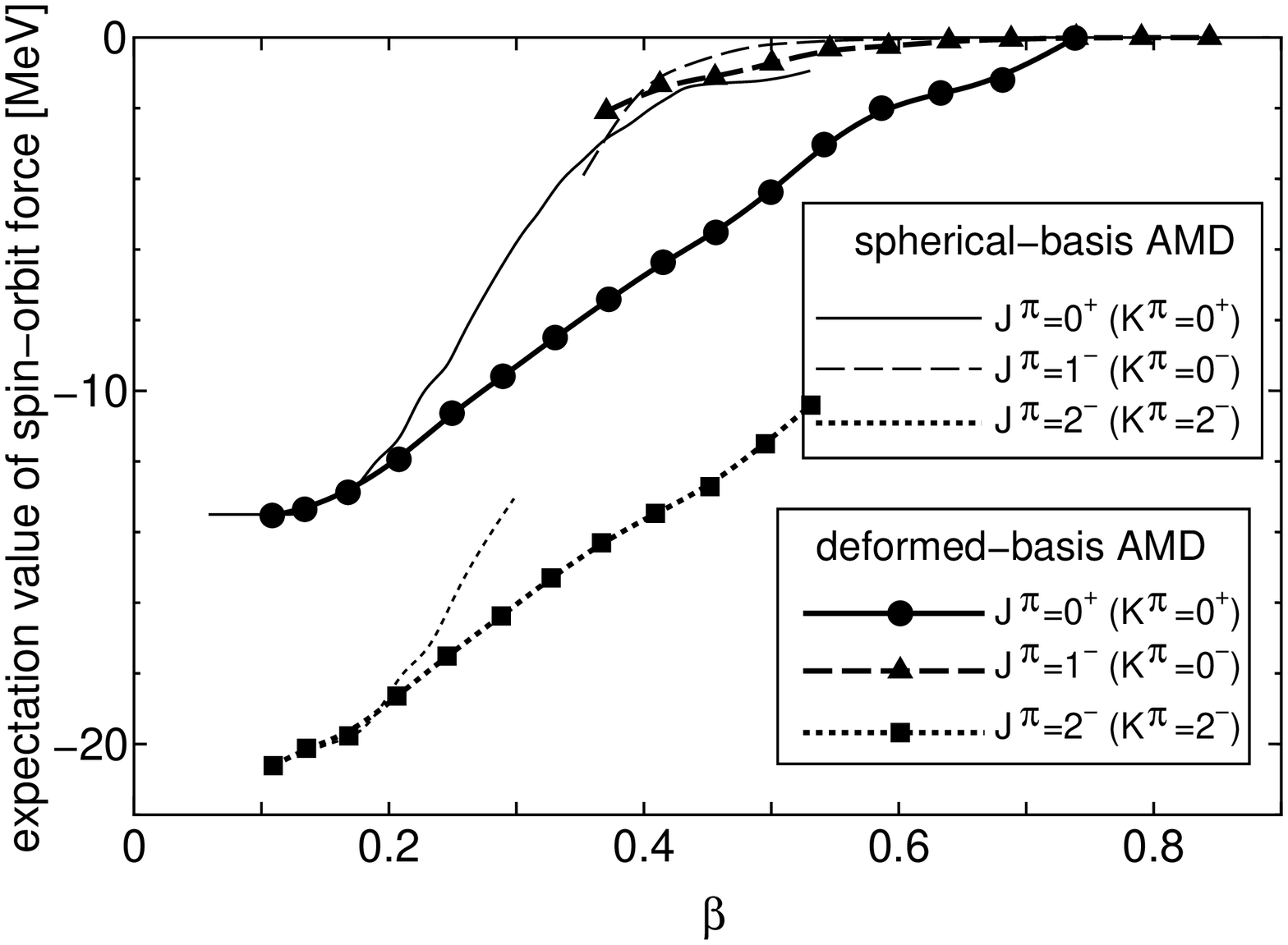}
\caption{The expectation value of the spin-orbit force for each parity
 projected wave function $\Phi^{J\pm}(\beta)$ are plotted as functions
 of the deformation parameter  $\beta$. Lines with symbols  (Lines) are 
 the  result of the deformed-basis AMD (spherical-basis AMD).}  
 \label{FIG_LS}
\end{figure}

After the variational calculation and the angular momentum projection
to the $J^\pi$=$0^+$ ($K^\pi$=$0^+$), $1^-$ ($K^\pi$=$0^-$) and $2^-$
($K^\pi$=$2^-$), we obtain three energy curves as functions of the
deformation parameter $\beta$ (FIG. \ref{FIG_SURFACE}). In the
positive-parity state, we obtain the $J^\pi$=$0^+$ curve (solid line
with circles in FIG. \ref{FIG_SURFACE}), and when we perform the GCM
calculation, $J^\pi$=$0^+_1$ and $J^\pi$=$0^+_4$ states are mainly
obtained from this curve. Similarly, $J^\pi$=$2^+_1$ and $J^\pi$=$2^+_4$
states are obtained by superposing the wave functions on the $J^\pi$=$2^+$
($K^\pi$=$0^+$) curve and so on for other higher spin states. In the
negative-parity state, we obtain the $J^\pi$=$1^-$ curve (dashed line
with triangle) and $2^-$ curve (dotted line with boxes) which mainly
contribute to the $J^\pi$=$1^-$ ($K^\pi$=$0^-_1$) and $J^\pi$=$2^-$
($K^\pi$=$2^-$) states, respectively. 
Detailed results of the GCM calculation are given in the
next subsection, and in this subsection, we concentrate on the
$J^{\pi}$=$0^+$, $0^-$ and $2^-$ curves to discuss the different nature
of the wave functions. The results obtained by spherical-basis AMD 
are also plotted for comparison (lines without
marks). Three patterns of the nuclear shape change mentioned in the 
previous section can be seen on these three energy curves obtained by
 deformed-basis AMD. When we compare the $J^\pi$=$1^-$
($K^\pi$=$0^-$) curve of deformed-basis AMD with that of 
spherical-basis AMD, they are almost identical, while there are
differences in the position of the minimum energy point and in the
minimum energy in the $J^{\pi}$=$2^-$ ($K^\pi$=$2^-$)
curves. In deformed-basis AMD, the $J^\pi$=$2^-$ curve has a minimum
energy point around $\beta$=$0.45$, but in spherical-basis AMD, it
has an unclear minimum energy point around 
$\beta$=$0.25$ and its energy is underestimated by about 5 MeV compared
with  deformed-basis AMD. These similarity and difference are due to the 
different nature of the $J^\pi$=$1^-$ and  $2^-$ curves.
The $J^\pi$=$1^-$ curve has an almost pure cluster structure, while 
the $J^\pi$=$2^-$ curve has a deformed mean-field structure. Since the
cluster structure is well described by both spherical- and deformed-basis
AMD, there is no difference in their description of the $J^\pi$=$1^-$
curve. However, because deformed-basis AMD can describe the deformed
mean-field structure better than spherical-basis AMD, there is
a difference in the description of the $J^\pi$=$2^-$ curve. The different
nature of these two states becomes clear in the deformation of the
single-particle wave packets and in the distributions of the centroids
of the single-particle wave packets. In the $J^\pi$=$1^-$ curve (dashed
line in FIG. \ref{FIG_WIDTH}), single-particle wave packets are always
almost spherical but the centroids of the twenty single-particle wave
packets (black points in FIG. \ref{FIG_Z} (d) and (e)) are widely
separated into two subunits, $\alpha$ (four points degenerate on
the upper side) and $^{16}{\rm O}$ (sixteen points are almost
degenerate in the middle). As the nuclear deformation   
becomes larger, only the distance between two subunits becomes
larger. Therefore, this nuclear shape change corresponds to the second
pattern in FIG. \ref{FIG_ILLUST}. In contrast, in the $J^\pi$=$2^-$
curve, the single-particle wave packets are significantly deformed, as
a function of the nuclear deformation (dotted line in
FIG. \ref{FIG_WIDTH}). All of the centroids of the twenty
single-particle wave packets are gathered around the center of the
nucleus (FIG. \ref{FIG_Z} (f) and  (g)), even though the nuclear
deformation becomes larger. We note that though the nuclear deformation
is larger in wave function (g) than in  wave function (d), the
centers of the single-particle wave packets are crowded into the center
of the nucleus in wave function (g). Therefore, this state corresponds
to the first pattern in FIG. \ref{FIG_ILLUST} and shows a deformed
mean-field nature. These different characters of the $J^\pi$=$1^-$ and
$2^-$ curves are directly reflected in the expectation value of the
spin-orbit force (FIG. \ref{FIG_LS}). The spin-orbit force does not act
at all in this state (dashed line with triangle in
FIG. \ref{FIG_LS}). However, it acts strongly in the $J^\pi$=$2^-$ wave
function (dotted line with boxes) which has an $(0p)^{-1}(sd)^5$
structure. The $(0p)^{-1}(sd)^5$ structure of the calculated
$J^\pi$=$2^-$ state is confirmed by investigating the structure of the
single-particle orbit. The technique for studying a single-particle orbit
structure within the AMD framework was developed by D\'ote, 
et al \cite{DOTE}.  

The energy curve for the $J^\pi$=$0^+$ state is more deeply bound about
5 MeV at the minimum energy point in  deformed-basis AMD than in 
spherical-basis AMD. The intermediate character of the $J^\pi$=$0^+$
curve is confirmed in the deformation and distribution of 
single-particle wave packets. In this curve, the deformation of
single-particle wave packets (solid line in FIG. \ref{FIG_WIDTH})
becomes larger as the nuclear deformation 
becomes larger in the small- and medium-deformation region
($\beta<0.5$). In this stage, the centroids of the single-particle wave
packets are gathered around the center of the nucleus
(FIG. \ref{FIG_Z}(a)) and the nucleus has a deformed mean-field
character. But the deformation of the single-particle wave packets is
saturated around $\beta\sim 0.5$. At this stage, the centroids of
the single-particle wave packets are separated into $\alpha$ and
$^{16}{\rm O}$ parts (FIG. \ref{FIG_Z}(b)), though the distance  
between them is rather small (about 1fm). Therefore, the nucleus has a
good amount of $\alpha$+$^{16}{\rm O}$ cluster component,
although the cluster structure is distorted by the deformed mean-field
structure. After this stage, the distance between the two subunits
becomes larger, as the nuclear deformation becomes larger, while the
deformation of the single-particle wave packets does not change. Then
around $\beta\sim0.65$, the distortion eases and the single-particle wave
packets rapidly become spherical. Thus, the $J^\pi$=$0^+$ curve
corresponds to the third pattern in FIG. \ref{FIG_ILLUST}. 
This change of nuclear shape and structure as a function
of the deformation parameter $\beta$ takes place gradually in the case
of deformed-basis AMD. Therefore, the expectation value of the
spin-orbit force decreases uniformly in deformed-basis AMD, while
in spherical-basis AMD it becomes zero rather rapidly.

\subsection{Low-lying level structure in $^{\bf 20}{\bf Ne}$}
\begin{figure}
\includegraphics[width=1.35\hsize]{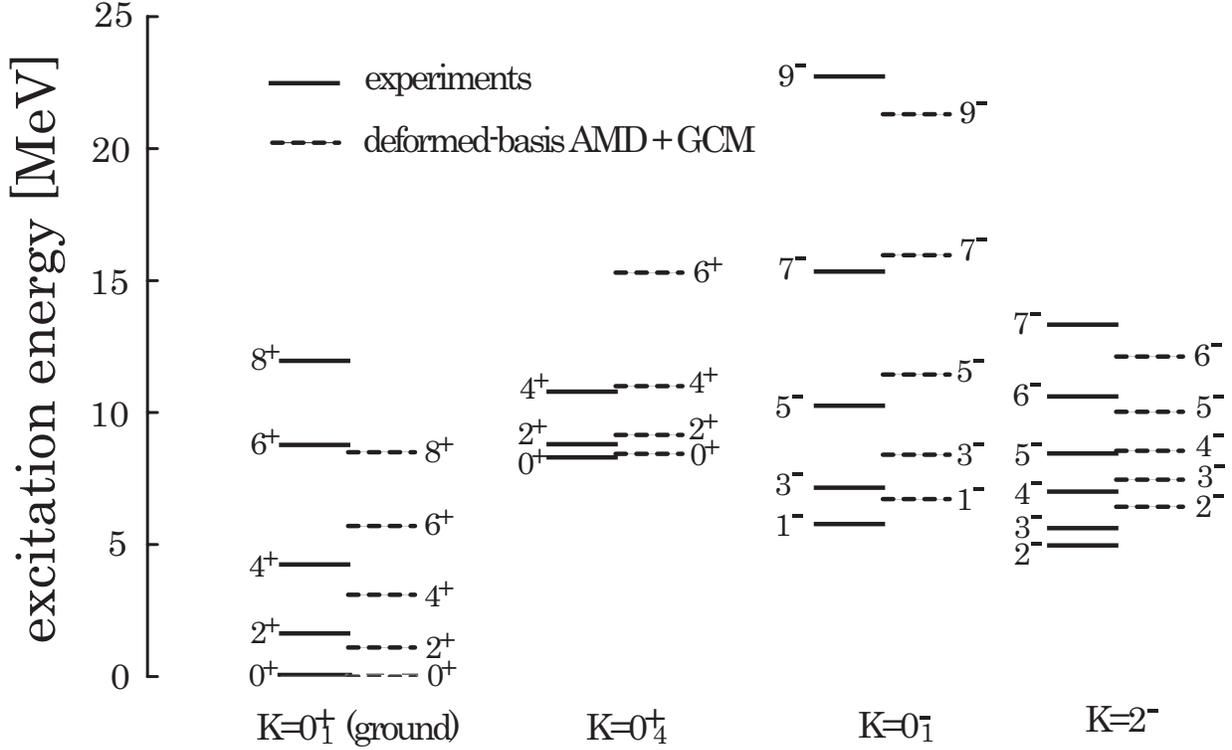}
\caption{The excitation energies of the low-lying states of $^{20}{\rm Ne}$. The observed values (the results of the deformed-basis AMD+GCM calculation) are given by the solid lines (dashed lines). The spin-parity of each state is also shown.}
\label{FIG_LEVEL}
\end{figure}

After the angular momentum projection, we perform the GCM calculation by
superposing all of the wave functions on the obtained energy curves, and
calculate the level scheme of $^{20}{\rm Ne}$ (FIG. \ref{FIG_LEVEL}). We 
have obtained two $K^\pi$=$0^+$ bands in positive-parity and
$K^\pi$=$2^-$ and $K^\pi$=$0^-_1$ bands in negative-parity. 
In this subsection, we discuss the detailed character of
these bands and show the related quantities in due order.

\subsubsection{$K^\pi=0^-_1$ band}
\begin{table}
 \caption{Observed and calculated $\alpha$-RW ($\theta_{\alpha}^2$),
 multiplied by 100 at the channel radius $a=6$fm, for $K^\pi$=$0^+_1$,
 $0^+_4$ and $0^-_1$ band members. For comparison, the
 results of the $(sd)^4$ shell model (SM) \cite{SDSM},
 $\alpha$+$^{16}{\rm O}$ RGM (RGM) \cite{AORGM} and
 $(\alpha$+$^{16}{\rm O})$+$(^{8}{\rm Be}+^{12}{\rm C})$  coupled
 channel OCM (OCM) \cite {CCOCM} are shown.}\label{TAB_A_WIDTH}  
\begin{ruledtabular}
\begin{tabular}{l|c|c|c|c|c|c}
$K^\pi$ & $J^\pi$ & $\theta_{obs}(a)^2\times 100$ &SM&
RGM & OCM & AMD\\\hline
$0^+_1$ & $6^+_1$ & 1.0$\pm$0.2 & 0.20& 1.4& 0.50& 0.53\\
        & $8^+_1$ & 0.094$\pm$0.027 & 0.020 & 0.24 & 0.10 &0.08\\\hline
$0^-_1$ & $1^-$ & $>$16 & & 34 & 22.2 & 31.0\\
      & $3^-_2$ &  26   & & 34 & 15.0 & 29.1\\
      & $5^-_2$ & 30    & & 36 & 11.5 & 28.8\\
      & $7^-_2$ & 22$\pm$5& & 33 & 12.5 & 11.5\\
      & $9^-$ & 17    & & 20 & 3.6 & 8.9\\\hline
$0^+_4$ & $0^+_4$& $>$50  & & & 139 & 69.0\\
 & $2^+_4$& $>$59  & & & 80 & 68.0\\
 & $4^+_4$ & 23 & & & 115 &   35.5\\
\end{tabular}
\end{ruledtabular}
\end{table}
\begin{table}
 \caption{The squared amplitude of the $\alpha$+$^{16}{\rm O}$ component
 $W^J$ (see appendix \ref{app::RWA}) and the  expectation  value  of the
 spin-orbit force   $\langle \hat{V}_{ls}\rangle$ (in MeV) of the GCM
 wave function for each state.}\label{TAB_OVERLAP_CLUSTER}    
\begin{ruledtabular}
\begin{tabular}{c|c|c|c||c|c|c|c}
$K^\pi$ & $J^\pi$ & $W^J$ &$\langle
 \hat{V}_{ls} \rangle$ & $K^\pi$ & $J^\pi$ & $W^J$ & $\langle
 \hat{V}_{ls} \rangle$\\\hline  
$0^+_1$ & $0^+_1$ & 0.70 & -5.2 & $0^-_1$ & $1^-_1$ & 0.95 & -0.8 \\
& $2^+_1$ & 0.68 & -5.3 &  & $3^-_2$ & 0.93 & -0.8 \\
& $4^+_1$ & 0.54 & -5.9 &  & $5^-_2$ & 0.88 &  -0.7 \\
& $6^+_1$ & 0.34 & -8.4 &  & $7^-_2$ & 0.71 &  -0.9\\
& $8^+_1$ & 0.28 & -10.9 &  & $9^-_2$ & 0.70 &  -1.3\\\hline
$0^+_4$ & $0^+_4$ & 0.82 (0.71) & -3.2 & $2^-$ & $2^-_1$ &  &  -12.9  \\
& $2^+_4$ & 0.81 (0.71) & -3.0 & & $3^-_1$ &  &  -13.0\\
& $4^+_4$ & 0.79 (0.57) & -4.9 & & $4^-_1$ &  &  -14.1\\
& $6^+_4$ & 0.67 (0.37) & -6.8 & & $5^-_1$ &  &  -14.4\\
& $8^+_4$ & 0.55 (0.38) & -7.4 & & $6^-_1$ &  &  -16.5\\
\end{tabular}
\end{ruledtabular}
\end{table}
Experimentally, the $K^\pi$=$0^-_1$ band is built upon the $1^-$ state
at 5.78 MeV, that is 1.05 MeV above the $\alpha$+$^{16}{\rm O}$
threshold. In the present calculation, the excitation energy of the
$1^-$ state is 6.72 MeV and the excitation energies of the higher spin
states show good agreement with the observed values. 

The $K^\pi$=$0^-_1$ band is regarded as having an almost pure
$\alpha$+$^{16}{\rm O}$ cluster structure, since the band members have
large $\alpha$ decay widths comparable with the Wigner limit. Indeed, in
the present calculation, the obtained $\alpha$ decay widths are 
large enough to be comparable with the observed values
(Table. \ref{TAB_A_WIDTH}) and the wave functions of the member states
have quite large amounts of the $\alpha$+$^{16}{\rm O}$ component
(Table. \ref{TAB_OVERLAP_CLUSTER}). The definitions and the method to 
evaluate these quantities are given in the Appendix. The almost pure
cluster structure of this band is due to a dominant contribution from
the wave functions on the $K^\pi$=$0^-$ ($J^\pi$=$1^-, 3^-,...$) energy
curves. As is discussed in the previous subsection, the wave  
functions on the $J^\pi$=$0^-$ energy curve have an
almost pure $\alpha$+$^{16}{\rm O}$ cluster structure (and this
character is common to other $J^\pi$ curves with $K^\pi$=$0^-$).
As a consequence, the member states of this band have an almost pure 
$\alpha$+$^{16}{\rm O}$ cluster structure and small expectation
values of the spin-orbit force
(Table. \ref{TAB_OVERLAP_CLUSTER}). Though we did not make any  
assumptions about the cluster structure in our calculation, the band
members have almost pure cluster structures. This fact means that the
cluster structure can exist as an independent degree of freedom of 
nuclear excitation, even though the effects of the spin-orbit force and
the deformation of the mean-field are taken into account in the model
space.  

\begin{figure}
\includegraphics[width=1.0\hsize]{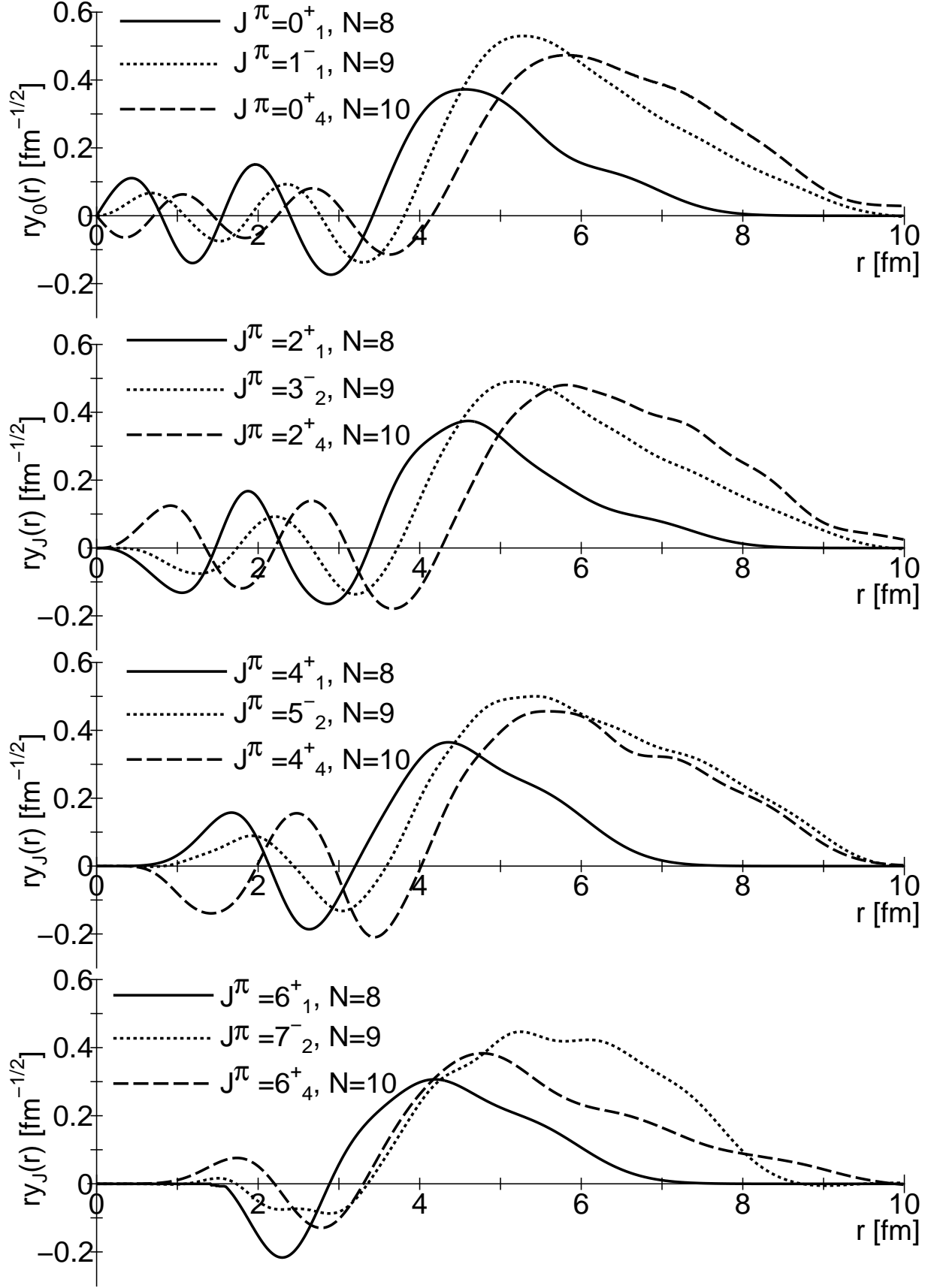}
\caption{The $\alpha$-RWA for the member states of the $K^\pi$=$0^+_1$
 (solid lines), $0^+_4$ (dashed lines) and $0^-_1$ (dotted lines)
 bands. See text and the appendix \ref{app::RWA}.}
\label{FIG_RWA}
\end{figure}

The relative motions between $\alpha$ and $^{16}{\rm O}$ clusters were
also investigated following the traditional cluster model studies. The
$\alpha$ reduced width amplitudes ${\cal Y}_L(a)$ ($\alpha$-RWA) are
shown in Fig. \ref{FIG_RWA} (dotted lines). 
The definitions and the calculational procedure for the $\alpha$-RWA are
also presented in the appendix \ref{app::RWA}. As in other cluster studies, 
it is clear that $K^\pi$=$0^-_1$ band members belong to the
$N=2n+L=9$ band members of the $\alpha$+$^{16}{\rm O}$ cluster
structure. Here, $N$ denotes the principal quantum number of 
relative motion and $n$ denotes the radial quantum number (number of
nodes). We also see that relative motion is suppressed inside 
of the nucleus and enhanced outside, which induces large
intra-band $E2$ transition probabilities 
(Table. \ref{TAB_E2_0}). 
\begin{table}
\caption{Observed (EXP) and calculated (AMD) intra-band $E2$ transition
probabilities B($E2$;$J_i^\pi\rightarrow J_f^\pi$) [${\rm e}^2{\rm fm}^4$]
within the   $K^\pi$=$0^-_1$ band. For comparison, $\alpha$-$^{16}{\rm O}$ RGM (RGM) \cite{AORGM} and multicluster GCM (5$\alpha$)\cite{BAYE} are also shown. The effective charges which are used  in each calculation are given in the 
bottom line.} \label{TAB_E2_0} 
\begin{ruledtabular}
\begin{tabular}{l|c|c|c|c|}
$J^\pi_i\rightarrow J^\pi_f$& EXP &  RGM & 5$\alpha$GCM & AMD\\\hline 
$3^- \rightarrow 1^-$& $164\pm26$ & 121.4 & 155 &151.2\\
$5^- \rightarrow 3^-$&  &133.399  & 206  & 182.4\\
$7^- \rightarrow 5^-$&  & 122.74 &  &141.6\\
$9^- \rightarrow 7^-$&  & 67.83 &  &87.9\\
\hline
$\delta e/e$  & & 0 & 0 & 0
\end{tabular}
\end{ruledtabular}
\end{table}

\subsubsection{Ground band ($K^\pi=0^+_1$ band)}
The obtained binding energy of the ground state is -159.8 MeV, which
underestimates the observed value by about 0.8 MeV. The rotational
character of the $0^+_1$, $2^+_1$ and $4^+_1$ states and the deviation
from the rotational spectra of the $6^+_1$ and $8^+_1$ states are
reproduced, though the obtained excitation energies are smaller than
the observed values.

The ground band must be the parity doublet partner of the
$K^\pi$=$0^-_1$ band, since the ground band is the only $K^\pi$=$0^+$
band below the $K^\pi$=$0^-_1$ band. However, at the same 
time, the existence of the deformed mean-field component in the ground
band has been discussed. For example, if the pure $\alpha$+$^{16}{\rm
O}$ cluster structure of this band is assumed, the intra-band $E2$ 
transition probabilities are underestimated. Shell model
calculation has shown the existence of broken symmetry components
in the wave function that activates the spin-orbit force.  

\begin{figure}
\includegraphics[width=\hsize]{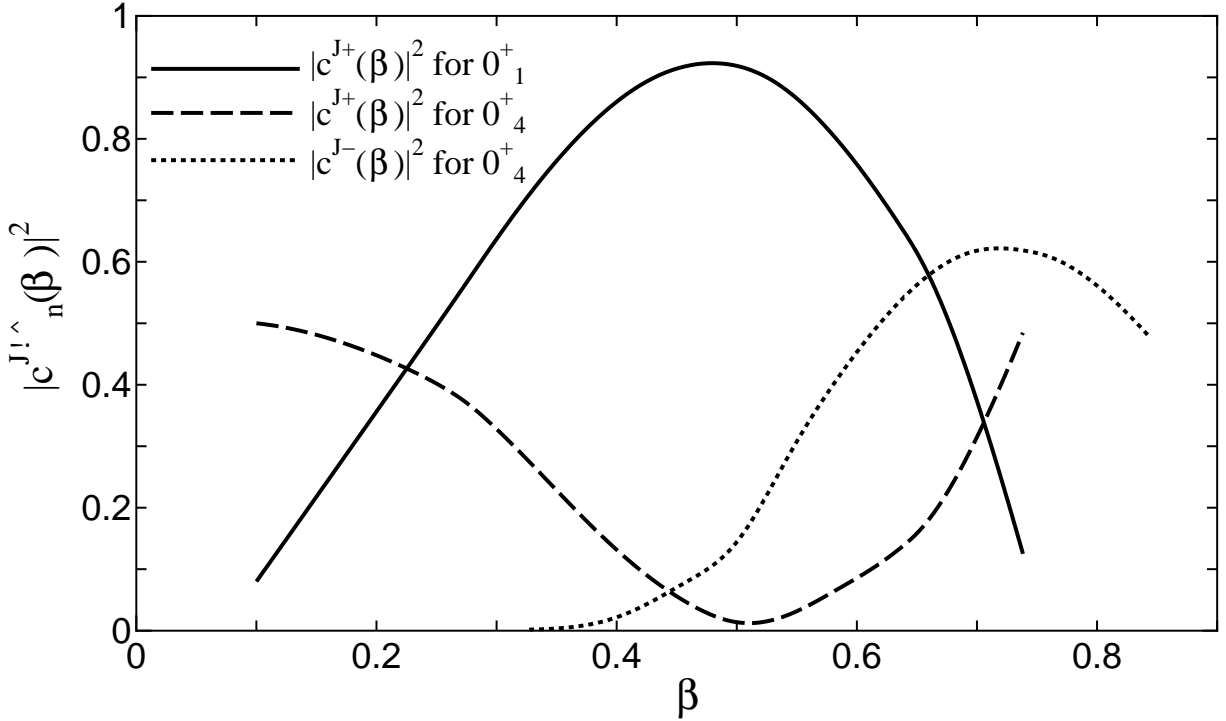}
\caption{The GCM amplitudes for $0^+_1$ (solid line) and $0^+_4$ (dashed
 line and dotted line) states. The definition of the GCM amplitude is
 given in the text.}\label{FIG_GCM_AMP}
\end{figure}

The GCM wave functions describing the member states of this band have
a mixed character embodying both the cluster structure and the deformed
mean-field structure. They have a dominant contribution from the wave
functions on the $K^\pi$=$0^+$ ($J^\pi$=$0^+, 2^+,...$) curves. The
GCM-amplitude of the $0^+_1$ state is shown in FIG. \ref{FIG_GCM_AMP}
(solid line). The GCM-amplitude 
that serves as a kind of measurement of the contribution to the GCM wave 
function from each intrinsic wave function is defined as 
\begin{eqnarray}
 |c^{J\pm}_n(\beta_0)|^2 \equiv |\langle \Phi_n^{J\pm}|\Phi^{J\pm}_{MK}(\beta_0)\rangle |^2,
\end{eqnarray}
where $\Phi_n^{J\pm}$ is the GCM wave function (Eq.\ref{EQ_GCM_WF}) and
$\Phi^{J\pm}_{MK}(\beta_0)$ is the parity and angular momentum projected
wave function (Eq.\ref{EQ_ANGULAR_WF}).
The GCM-amplitude reaches a maximum
around $\beta\sim 0.50$ in the low spin state and is larger than 50\%
from $\beta\sim 0.20$ to $\beta\sim 0.70$. As is shown in the previous
subsection, the intrinsic wave function on the $J^\pi$=$0^+$ curve
starts to change its structure from a deformed mean-field-like 
structure to a cluster structure around $\beta$=$0.5$. 
Therefore, unlike the $K^\pi$=$0^-_1$ band, the $K^\pi$=$0^+_1$ band is
a mixture of the cluster structure and the deformed mean-field
structure. Indeed, the wave functions of the ground band have smaller
overlap with the  $\alpha$+$^{16}{\rm O}$ model space than the
$K^\pi$=$0^-_1$ band (Table.\ref{TAB_OVERLAP_CLUSTER}). This transient
nature of the ground band is reflected in the expectation value of the
spin-orbit force, which amounts to about -5.0 MeV in the ground state
and increases as the angular momentum increases. The increase of the
spin-orbit force contribution is a consequence of the structure change
along the yrast line or the anti-stretching phenomenon, which is
discussed later. 

\begin{table}
\caption{Observed (EXP) and calculated (AMD) intra-band $E2$ transition
probabilities B($E2$;$J_i^\pi\rightarrow J_f^\pi$) [${\rm e}^2{\rm fm}^4$]
within the   $K^\pi$=$0^+_1$ band. For comparison, the results of the
 $(sd)^4$ shell model (SM) \cite{ARIMA}, 
 $\alpha$+$^{16}{\rm O}$ RGM (RGM) \cite{AORGM},  
($\alpha$+$^{16}{\rm O}$)+($^{8}{\rm Be}$+$^{12}{\rm C}$)
coupled channel OCM (OCM) \cite{CCOCM} and multicluster GCM (5$\alpha$GCM)
are also shown. The effective charges which were used in each
calculation are given in the bottom line.} \label{TAB_E2_1} 
\begin{ruledtabular}
\begin{tabular}{l|c|c|c|c|c|c}
$J^\pi_i\rightarrow J^\pi_f$& EXP &
 SM&  RGM & OCM &5$\alpha$GCM&AMD\\\hline 
$2^+_1 \rightarrow 0^+_1$& $65\pm 3$ &57.0  & 36.2 &57.0&50.0 &70.3\\
$4^+_1 \rightarrow 2^+_1$& $71\pm 6$ &69.9 & 45.22 & 70.9 &64.4 &83.7\\
$6^+_1 \rightarrow 4^+_1$& $64\pm 10$ &57.9 &36.5 & 57.1 &55.3 & 52.7\\
$8^+_1 \rightarrow 6^+_1$& $29\pm 4$ &35.5  & 19.7& 34.8 & &21.0\\
\hline
$\delta e/e$ &  &0.54 & 0 & 0.155 & 0&0
\end{tabular}
\end{ruledtabular}
\end{table}

The transient nature of the ground band affects the intra-band
$E2$ transition probabilities (Table.\ref{TAB_E2_1}). 
They are successfully described quantitatively by the cluster models
(RGM \cite{AORGM} and coupled channel OCM \cite{CCOCM}), 
but they need to introduce a small effective charge. When the model
space is expanded to include the $\alpha$+$^{16}{\rm O}(0^+, 3^-, 1^-)$
channels (multicluster GCM)\cite{BAYE}, the deformation of the
mean-field is described to some extent and the result is improved. In
the present calculation, the $E2$ transition probabilities of the ground
band are described well without any effective charges. Therefore, we can
conclude that the general trend of the intra-band $E2$ transition
probabilities are well described qualitatively by the
$\alpha$+$^{16}{\rm O}$ cluster  component in the wave functions and the
existence of the deformed mean-field  component (or the cluster
dissociated component) slightly  enlarges the intra-band $E2$ transition
probabilities.     

Another important feature of the $\alpha$+$^{16}{\rm O}$ cluster
structure is the anti-stretching phenomenon, which is studied by many
cluster models \cite{P_SHELL_CLUSTERS}. As the angular momentum
increases, the $\alpha$+$^{16}{\rm O}$ cluster structure moderates and
the average 
distance between $\alpha$ and $^{16}{\rm O}$ becomes smaller. Indeed, in
the present result, the position of the maximum peak of
the $\alpha$-RWA moves slightly toward the inside of the nucleus as the
angular momentum increases, though the tendency is rather mild. However,
we wish to emphasize the difference of the anti-stretching phenomenon
described by the $\alpha$+$^{16}{\rm O}$ cluster models and the present
model. In the $\alpha$+$^{16}{\rm O}$ cluster models, the
anti-stretching phenomenon is described by the decreasing distance
between $\alpha$ and $^{16}{\rm O}$. However, in the present
model, as the angular momentum increases, the $\alpha$+$^{16}{\rm O}$
component diminishes and the spin-orbit force acts
strongly. In particular, the $8^+_1$ state is 
oblately deformed and four valence nucleons are aligned, which leads to
a large contribution of the spin-orbit force which amounts to about -10
MeV, though a small amount of the $\alpha$+$^{16}{\rm O}$ component
survives outside of the nucleus. We also note that this remaining
small amount of $\alpha$+$^{16}{\rm O}$ component outside of the
nucleus leads to the relatively large $\alpha$ decay widths of the
$J^\pi$=$6^+_1$ and $8^+_1$  states (Table. \ref{TAB_A_WIDTH}).

\subsubsection{$K^\pi=0^+_4$ band}
The $K^\pi$=$0^+_4$ band has a 
prominent $\alpha$+$^{16}{\rm O}$ cluster structure in which the
relative motion between clusters is $2\hbar\omega$ excited
($N=2n+L=10$), while $\alpha$ and $^{16}{\rm O}$ clusters are not
excited. In the present calculation,
the obtained second positive-parity band members have large amounts of
$\alpha$+$^{16}{\rm O}$ component and large $\alpha$ decay widths. Also,
the relative motion between $\alpha$ and $^{16}{\rm O}$  has the
principal quantum number $N=10$. Therefore, we identify this band as the 
$K^\pi$=$0^+_4$ band. The $K^\pi$=$0^+_2$ and $0^+_3$ bands which do
not have the $\alpha$+$^{16}{\rm O}$ structure are not
obtained in the present calculation, but we consider that we will obtain
these two bands when we make the variational calculation and the GCM
calculation in the model space which is orthogonal to the $K^\pi$=$0^+$
($J^\pi$=$0^+, 2^+,...$) curves obtained in the present calculation. The
calculated and observed excitation energy of the $0^+_4$ state are 8.44
MeV and 8.3 MeV, respectively. 

We consider it worthwhile to discuss how the prominent
$\alpha$+$^{16}{\rm O}$ cluster structure of this band compared to the
ground band is derived in the present calculation. As explained in
section \ref{SEC_FRAME}, we include all of the obtained intrinsic wave
functions as the GCM basis. The prominent $\alpha$+$^{16}{\rm O}$ cluster
structure is due to the considerable contribution of the
$\Phi_{int(-)}(\beta)$ which was obtained by the variational
calculation after the projection to negative-parity. When we include
only $P^{J^+}_{MK=0}\Phi_{int(+)}(\beta)$ in the GCM basis, the content
of $\alpha$+$^{16}{\rm O}$ components of $K^\pi$=$0^+_4$ band members is
around 40$\sim$70\% (bracketed numbers in the third column of Table
\ref{TAB_OVERLAP_CLUSTER}) and comparable with that of the 
ground band. In contrast, when we include the
$P^{J^+}_{MK=0}\Phi_{int(-)}(\beta)$  
as well as  $P^{J+}_{MK=0}\Phi_{int(+)}(\beta)$, the
$\alpha$+$^{16}{\rm O}$ cluster structure of this band becomes prominent
(numbers in the third column of Table \ref{TAB_OVERLAP_CLUSTER}). 
In the outside of nucleus, the contribution from
$P^{J^+}_{MK=0}\Phi_{int(-)}(\beta)$ (dotted line in
Fig. \ref{FIG_GCM_AMP}) becomes larger, while in the inside
$P^{J^+}_{MK=0}\Phi_{int(+)}(\beta)$ (dashed line in
Fig. \ref{FIG_GCM_AMP}) is dominant. Therefore, in the outside of
nucleus, the $\alpha$+$^{16}{\rm  
O}$ cluster structure is prominent, which leads to a large $\alpha$
decay width. However, in the inside, the mean-field-like structure
contributes to some extent and it leads to the existence of the small
non-$\alpha$+$^{16}{\rm O}$ component. It is interesting that though 
$P^+_{MK=0}\Phi_{int(-)}(\beta)$ is less bound than
$P^+_{MK=0}\Phi_{int(+)}(\beta)$, $P^+_{MK=0}\Phi_{int(-)}(\beta)$
contributes to this band.

\subsubsection{$K^\pi=2^-$ band}
\begin{table}
\caption{Observed and calculated intra-band $E2$ transition
 probabilities B($E2$;$J^\pi_i\rightarrow J^\pi_f$) [${\rm e}^2{\rm
 fm}^4$] within the $K^\pi$=$2^-$ band. For comparison, the results of
 the $j$-$j$ shell model (SM) \cite{JJSM} and 
 ($\alpha$+$^{16}{\rm O}$)+($^{8}{\rm Be}$+$^{12}{\rm C}$)
 coupled channel OCM (OCM) \cite{CCOCM} are also shown. The effective
 charges which were used in each calculation are given in the bottom
 line.} \label{TAB_E2_2} 
\begin{ruledtabular}
\begin{tabular}{c|c|c|c|c}
$J^\pi_i\rightarrow J^\pi_f$& EXP&
 SM& OCM & AMD+GCM\\\hline
$3^-_1 \rightarrow 2^-$& $113\pm29$ & 97  & 108 & 102.8\\
$4^- \rightarrow 3^-_1$& $77\pm16$ & 75 & 77 & 77.8\\
$4^- \rightarrow 2^-$& $34\pm6$ & 36 & 34 & 38.5\\
$5^-_1 \rightarrow 4^-$& $<808$ & 44 & 45 & 84.5\\
$5^-_1 \rightarrow 3^-_1$& $84\pm19$ &48  & 49 & 56.6\\
$6^- \rightarrow 5^-_1$& $32\pm13$ & 32 & 34 & 29.9\\
$6^- \rightarrow 4^-$& $55^{+23}_{-13}$ & 51 & 67 & 64.0\\\hline
$\delta e/e$ &  & 0.8 & 0.069 & 0
\end{tabular}
\end{ruledtabular}
\end{table}

The $K^\pi$=$2^-$ band is the lowest negative-parity band. It starts
from the $2^-$ state at 4.97 MeV and is observed up to the $9^-$ state,
and exhibits a good rotational spectrum. The obtained excitation energy
of the $2^-$ state is 6.51 MeV and we have calculated up to the $6^-$
state. 

This band has a quite different character from the other three bands.
From the shell model calculations, it is known that this band has a 
$(1p)^{-1}(sd)^5$ structure with respect to the $^{16}{\rm O}$ core and
the $SU_3$(8,2) configuration is dominant, which means the absence of the
$\alpha$+$^{16}{\rm O}$ cluster structure and a large contribution
from the spin-orbit force. Indeed, in the present result, 
the GCM wave functions of this band members consist of the wave
functions on the $K^\pi$=$2^-$ curve and in the natural parity states
($\pi=(-)^J$), the mixing with the wave functions on the $K^\pi$=$0^-$
curve is quite small. As discussed in the previous subsection, the wave
functions on the $K^\pi$=$2^-$ curve have a deformed mean-field-like
character. Therefore, the spherical-basis AMD gives much
higher excitation energy of the $2^-$ state, 11.3 MeV. 
Though the character of the band is quite different from that of
the $K^\pi$=$0^{\pm}_1$ band, the B($E2$) transition probabilities are
again described well without any effective charges. 

\section{Summary}
In this study, we have presented the details of the new theoretical
framework of deformed-basis AMD. This framework enables us to describe 
the coexistence, competition and mixing of the deformed mean-field
structure and the cluster structure and gives us the proper evaluation
of the spin-orbit force. 

We have applied this framework to the $^{20}{\rm Ne}$ nucleus. On the
analysis of the obtained energy curves, three different characteristics
of the nuclear structure appeared. The $J^\pi$=$1^-$ ($K^\pi$=$0^-$)
curve has an almost pure $\alpha$+$^{16}{\rm O}$ cluster structure, in
which the spherical single-particle wave packets are localized into two
parts. In contrast, the $J^\pi$=$2^-$ ($K^\pi$=$2^-$) curve has a
deformed mean-field-like character in which the significantly deformed
single particle wave packets are gathered around the center of the nucleus.
The $J^\pi$=$0^+$ ($K^\pi$=$0^+$) curve has a mixed
character. As the nuclear deformation becomes larger, the wave function
gradually changes its structure from a deformed mean-field-like
structure to a cluster structure. 

 By superposing the obtained wave functions, four rotational bands are
obtained in the low-energy region. The $K^\pi$=$0^-_1$ band 
has an almost pure $\alpha$+$^{16}{\rm O}$ cluster structure. Observed
large $\alpha$ decay widths of this band members are reproduced and the
obtained $\alpha$-RWAs show a quite similar trend to those obtained by 
other cluster models. These facts mean that the cluster structure can
exist as an independent degree of freedom of the nuclear excitation, even
though the deformation of the mean-field and the effects of the
spin-orbit force are included in the model space. The ground band
has a mixed character of the deformed mean-field structure and the cluster
structure. Because of the mixture of the deformed mean-field character,
the obtained intra-band $E2$ transition probabilities are enlarged
compared to the cluster models and the observed values are reproduced
without any effective charge. The structure change along the yrast line
is also investigated. As the angular momentum becomes larger, the
deformed mean-field component in the wave function increases and
the spin-orbit force acts strongly.  By including the wave functions
which have an almost pure $\alpha$+$^{16}{\rm O}$ 
structure as the GCM basis, the $K^\pi$=$0^+_4$ band acquires a
prominent cluster structure and observed large $\alpha$ decay widths are
reproduced. The $K^\pi$=$2^-$ band has a different
character from the other three bands. The wave functions of this band
member have a deformed mean-field character. The 
intra-band $E2$ transition probabilities are again reproduced without
any effective charge. 

Though these four bands have different characteristics, 
deformed-basis AMD successfully describes the coexistence and mixture of
the cluster structure and mean-field structure in the same nucleus.
Therefore, we believe that deformed-basis AMD is one of the
most powerful approaches to describe the interplay of the mean-field
structure and the cluster structure and to study the cluster structure
and exotic cluster-like structure in medium-heavy and neutron-rich
nuclei.

\begin{acknowledgments}
The author would like to thank Professor H. Horiuchi for his
encouragement and for many valuable discussions. He also thanks
 Dr. A. D\'ote,  Dr. Y. Kanada-En'yo and Dr. Y. Fujiwara for valuable
 discussions.    Many of the computations were carried out
 by SX-5 at the Research Center for Nuclear Physics, Osaka University
 (RCNP).  This work was partially performed under the Research Project
 for Study of Unstable Nuclei from Nuclear Cluster Aspect sponsored
 by the Institute of Physical and Chemical Research (RIKEN). 
\end{acknowledgments}

\appendix*
\section{reduced width amplitude of the AMD wave function} \label{app::RWA}
The calculational methods for the $\alpha$ reduced width amplitude
($\alpha$-RWA), $\alpha$ decay width and the square amplitude of the
$\alpha$+$^{16}{\rm O}$ component are briefly given. The method for
evaluating these quantities from the AMD wave function was developed by
Y. Kanada-En'yo \cite{ENYO_NEW} and the reader is directed to references
\cite{ENYO_NEW, P_SHELL_CLUSTERS} for more details.  

The $\alpha$+$^{16}{\rm O}$ system is generally expressed by the
RGM-type wave function. 
\begin{eqnarray}
 \Phi_{\alpha+^{16}{\rm O}}^{J^\pi} &=& n_0{\cal{A}}
  \{\chi_J(r)Y_{J0}(\hat{r})\phi(\alpha)\phi(^{16}{\rm O})\},\\
   \quad n_0 &=& \sqrt{16!\cdot4!/20!}, \quad \pi = (-)^J.
\end{eqnarray}
Here, $\cal{A}$ is the antisymmetrizer, ${\bf r}$ is the
relative coordinate between $\alpha$ and $^{16}{\rm O}$,
$\phi(\alpha)$ and $\phi(^{16}{\rm O})$ are the normalized internal wave
functions of the clusters. $\chi_J(r)$ is the radial wave function of the
relative motion between $\alpha$ and $^{16}{\rm O}$, and so normalized
that $\Phi_{\alpha+^{16}{\rm O}}^{J}$ is normalized to unity.

The normalized deformed-base AMD+GCM  wave function $\Phi_M^{J^\pi}$ of
$^{20}{\rm Ne}$ is divided into the $\alpha$+$^{16}{\rm O}$ component
and the residual part $\Phi_R^{J^\pi}$.
\begin{eqnarray}
 \Phi_M^{J^\pi} &=& \alpha\Phi^{J^\pi}_{\alpha+^{16}{\rm O}}
   + \sqrt{1-\alpha^2} \Phi_R^{J^\pi}.  \label{EQ_DECOMPOSE}
\end{eqnarray} 
$\Phi_{R}^{J^\pi}$ is also normalized and orthogonal to the
$\Phi_{\alpha+^{16}{\rm O}}^{J^\pi}$. 
We introduce the projection operator ${\cal P}_L$ which projects out the
$\alpha$+$^{16}{\rm O}$ component from the $\Phi_M^{J\pi}$,
\begin{eqnarray}
{\cal P}_L\Phi_M^{J\pi}&=&\alpha\Phi_{\alpha+^{16}{\rm O}}^{J\pi}\nonumber\\
                &=&\alpha n_0 {\cal A}\{\chi_J(r)Y_{J0}(\hat{r})\phi(\alpha)
		 \phi(^{16}{\rm O})\}.
  \label{EQ_PROJECTION} 
\end{eqnarray}
 The practical formula for ${\cal P}_L$ used in
this study is given later.
Using this projection operator, the squared amplitude of the
$\alpha$+$^{16}{\rm O}$ component $W_J$ of $\Phi_{M}^{J^\pi}$ is 
written as, 
\begin{eqnarray}
 W_J\equiv|\alpha|^2 = \langle \Phi^{J\pi}_M|{\cal P}_L|\Phi_M^{J\pi}\rangle.
  \label{EQ_AMOUNT}
\end{eqnarray}
The $\alpha$-RWA ${\cal Y}_J(a)$ is defined as 
\begin{eqnarray}
 {\cal Y}_J(a) &\equiv& \frac{1}{n_0} \langle \frac{\delta(r-a)}{r^2}
  Y_{J0}(\hat r) 
  \phi(\alpha)\phi(^{16}{\rm O})|\Phi^{J^\pi}_M\rangle \nonumber\\
 &=&\frac{\alpha}{n_0} \langle \frac{\delta(r-a)}{r^2}Y_{J0}(\hat r)
  \phi(\alpha)\phi(^{16}{\rm O})| \nonumber\\
&&\times|\Phi_{\alpha+^{16}{\rm O}}^{J^\pi}\rangle,
\end{eqnarray}
and its squared amplitude $|a{\cal Y}(a)|^2$ is the
probability to find $\alpha$ and $^{16}{\rm O}$ clusters at the
inter-cluster distance $r=a$.
 If we expand $\chi_J(r)$ with the radial
wave function of the H.O. with the width parameter
$\gamma$, $R_{NJ}(r,\gamma)$ as 
\begin{eqnarray}
 \chi_J(r) = \sum_N e_{NJ} R_{NJ}(r,\gamma), \label{EQ_HO_EXPAND}
\end{eqnarray}
the $\alpha$-RWA is written as 
\begin{eqnarray}
 {\cal Y}_J(a) = \alpha\sum_N \mu_N e_{NJ} R_{NJ}(a,\gamma). \label{EQ_RWA_EX}
\end{eqnarray}
This means that when $e_N$ is obtained, ${\cal Y}(a)$ is expanded by
the H.O. wave function.
Here, the eigenvalue of the RGM norm kernel $\mu_N$ is defined as 
\begin{eqnarray}
 \mu_N = &&\langle R_{nJ}(r)Y_{J0}({\hat r})\phi(\alpha)\phi(^{16}{\rm O})|
  \nonumber \\
 &&{\cal A}\{R_{nJ}(r)Y_{J0}({\hat r})\phi(\alpha)\phi(^{16}{\rm O})\}\rangle, 
\end{eqnarray}
and its value is analytically evaluated.
We calculate $\alpha$ decay width $\Gamma_\alpha$ using the $\alpha$-RWA.
\begin{eqnarray}
 \Gamma_\alpha &=& 2P_J(a)\gamma^2(a),\nonumber\\
 P_J(a) &=& \frac{ka}{F^2_L(ka)+G^2_L(ka)},\nonumber\\
\gamma^2(a) &=& \frac{\hbar^2}{2ma}|a{\cal Y}_L(a)|^2,
\end{eqnarray}
where $F_L$ and $G_L$ are the regular and irregular Coulomb
functions, $k$ is the wave number of the resonance energy, and $a$ is
the channel radius, which is chosen to be $a=6$fm.

To evaluate $W_J$ and ${\cal Y}_J(a)$, we have approximated the projection
operator ${\cal P}_L$ with the set of the orthonormalized
angular-momentum-projected Brink wave functions. We start from the Brink 
wave function $\varphi_B(R_i)$ in which $\alpha$ and $^{16}{\rm O}$ are
represented by the $SU_3$-limit wave functions which are 
located at the points $(0,0,\frac{16}{20}R_i)$ and
$(0,0,-\frac{4}{20}R_i)$, respectively,
\begin{eqnarray}
\varphi_B(R_i) &=& n_0{\cal A}\bigl\{\varphi(\alpha,\frac{16}{20}{\bf R}_i),
  \varphi(^{16}{\rm O},-\frac{4}{20}{\bf R}_i)\bigr\},\\
 {\bf R}_i &\equiv& (0,0,R_i).
\end{eqnarray}
The width parameters of $\alpha$ and $^{16}{\rm O}$ are taken to
be the same value $\nu$ for simplicity. By separating the center-of-mass
wave function $\omega({\bf X}_G)$, the internal wave function
$\varphi_C( R_i)$ is written as 
\begin{eqnarray}
\varphi_B(R_i)&=& \omega({\bf X}_G)\varphi_C(R_i),\\
\varphi_C(R_i)&=&n_0 {\cal A}\{\Gamma({\bf r},{\bf R}_i,\gamma)\phi(\alpha)  \phi(^{16}{\rm O})\},\label{EQ_BRINK_AJP}\\
\Gamma({\bf r},{\bf R}_i,\gamma)&=&\biggl(\frac{2\gamma}{\pi} \biggr)^{3/4}
e^{-\gamma({\bf r}-{\bf R}_i)^2} ,\label{EQ_BRINK_GAMMA}\\
\omega({\bf X}_G)&=&\biggl(\frac{20\cdot 2\nu}{\pi} \biggr)^{3/4}
\exp(-20\nu{\bf X}_G^2), \\
{\bf X}_G &=& \frac{1}{20}\sum_{i=1}^{20}{\bf r}_i,\quad
\gamma=\frac{16\cdot4}{20}\nu,
\end{eqnarray} 
where $\Gamma({\bf r},{\bf R_i},\gamma)$ is the relative wave function
between $\alpha$ and $^{16}{\rm O}$ of the Brink wave function.
The angular-momentum-projected Brink wave function $\phi^J_C(R_i)$ is
obtained from the $\phi_C(R_i)$,  
\begin{eqnarray}
 \varphi_C^J(R_i) = q_i^J P^J_{00}\varphi_C(R_i).\label{EQ_APROJ_BRINK}
\end{eqnarray}
$P^J_{00}$ is the angular momentum projector and $q_i^J$ is the
normalization factor.
The orthonormalized set of the angular-momentum-projected Brink wave
functions $\tilde{\varphi}_\alpha^J$ is described by the linear
combination of the  $\varphi_C^J(R_i)$, 
\begin{eqnarray}
 \tilde{\varphi}_\alpha^J &=& \frac{1}{\sqrt{\rho_\alpha}}
  \sum_{i}c_{i\alpha} \varphi_C^J(R_i),
\end{eqnarray}
and the coefficients $\rho_\alpha$ and $c_{i\alpha}$ are the eigenvalues
and eigenvectors of the overlap matrix $B_{ij}=\langle \varphi_C^J(R_i)|
\varphi_C^J(R_j)\rangle$,
\begin{eqnarray}
 B_{ij}c_{i\alpha} = \rho_\alpha c_{i\alpha}.
\end{eqnarray}
If enough numbers of the basis wave function
$\widetilde{\varphi}^{J}_\alpha$ are used, the projection
operator ${\cal P}_L$ is approximated by the set of 
$\widetilde{\varphi}^J_\alpha$,
\begin{eqnarray}
 {\cal P}_L &\simeq&
  \sum_{\alpha}|\widetilde{\varphi}_\alpha^J\rangle\langle\widetilde{\varphi}_\alpha^J|,\nonumber\\ 
  &=& \sum_{ij}B^{-1}_{ji}|\varphi_C^J(R_i)\rangle\langle\varphi_C^J(R_j)|.
   \label{EQ_PROJECTOR}
\end{eqnarray}
In the present calculation, we have employed 22 Brink-wave functions in
which $R_i$ is taken as $R_i=0.5, 1.0, 1.5,..., 11.0$ fm.

Using Eq. (\ref{EQ_PROJECTOR}), Eq. (\ref{EQ_PROJECTION}) and
Eq. (\ref{EQ_AMOUNT}) are rewritten as  
\begin{eqnarray}
{\cal P}_L\Phi_M^{J\pi} &=&\alpha n_0{\cal A}\{\chi_J(r)Y_{J0}(\hat{r})\phi(\alpha)\phi(^{16}{\rm O})\},\nonumber\\
 &\simeq&\sum_{ij}B_{ji}^{-1}\langle \varphi_C^J({R}_j)|\Phi_M^{J^\pi}\rangle 
   |\varphi_C^J(R_i)\rangle, \label{EQ_PROJECTION2}\\
|\alpha|^2 &\simeq& \sum_{ij}\langle \varphi_C^J({R}_j)|\Phi_M^{J^\pi}\rangle
 B^{-1}_{ji} \nonumber\\
&&\times\langle \Phi_M^{J^\pi}|\varphi_C^J({R}_i)\rangle.
\end{eqnarray}

The expansion coefficient $e_N$ of $\chi_J(r)$ and ${\cal Y}(a)$ is
obtained from Eq. (\ref{EQ_PROJECTION2}). We expand the wave function of the
relative motion $\gamma({\bf r}, {\bf R}_i, \gamma)$ of Eq. (\ref{EQ_GAM}),
\begin{eqnarray}
 \Gamma({\bf r},{\bf R_i},\gamma) = e^{-\frac{\gamma}{2}R_i^2}
  \sum_{NJm}\biggl\{\frac{(\gamma R_i^2)^\frac{N}{2}}{\sqrt{N!}}\label{EQ_GAM}\nonumber\\ 
\times\sqrt{\frac{4\pi}{2J+1}} 
  A^N_J Y_{Jm}({\hat R}_i)R_{NJ}(r,\gamma)Y_{Jm}(\hat r)\biggr\},\\
 A^N_J \equiv (-)^{(N-J)/2}\sqrt{\frac{(2J+1)\cdot N!}{(N-J)!!
  \cdot (N+J+1)!!}}.
\end{eqnarray}
And using Eq. (\ref{EQ_BRINK_AJP}) and (\ref{EQ_GAM}),
Eq. (\ref{EQ_APROJ_BRINK}) is rewritten as 
\begin{eqnarray}
 \varphi_C^J(R_i) &=& n_0{\cal A}\{\chi_J^{(i)}(r)Y_{J0}(\hat r)\phi(\alpha)
  \phi(^{16}{\rm O})\},\label{EQ_APROJ_BRINK2}\\
  \chi^{(i)}_J(r) &=& q_i^J e^{-\frac{\gamma}{2}R_i^2}\sum_{N}\frac{(\gamma R_i^2)^\frac{N}{2}}{\sqrt{N!}}A_J^NR_{NJ}(r).\label{EQ_BRINK_HO_EXPAND}
\end{eqnarray}
By inserting Eq. (\ref{EQ_APROJ_BRINK2}) into
Eq. (\ref{EQ_PROJECTION2}), we obtain $\chi_J(r)$ as a superposition of $\chi^{(i)}_J(r)$,
\begin{eqnarray}
\chi_J(r) \simeq \frac{1}{\alpha}\sum_{ij}\langle\varphi_C^J({R}_j)|\Phi_M^{J^\pi}\rangle B^{-1}_{ji}
  \chi^{(i)}_J(r).\label{EQ_ANS1}
\end{eqnarray}
Inserting Eq. (\ref{EQ_RWA_EX}) and (\ref{EQ_BRINK_HO_EXPAND}) into Eq. (\ref{EQ_ANS1}), $e_N$
is given as 
\begin{eqnarray}
  e_N &=& \frac{1}{\alpha}A^N_J\sum_{ij}\biggl\{\langle\varphi_C^{J}({R}_j)
  |\Phi_M^{J^\pi}\rangle B^{-1}_{ji}\nonumber\\
 &&\times q_i^J e^{-\frac{\gamma}{2}{ R}^2_i}
  \frac{(\gamma {R}^2_i)^{\frac{N}{2}}}{\sqrt{N!}}\biggr\}.
\end{eqnarray}

\bibliography{apssamp}
\end{document}